\documentclass[12pt]{article}
\usepackage[includefoot,margin=2.95cm]{geometry}
\usepackage{amssymb}
\usepackage{amsmath}
\usepackage{graphicx}
\usepackage{sectsty}
\usepackage{url}
\newcommand{\srr}{\raggedright}
\newcommand{\D}{\Delta}
\begin{document}
\allsectionsfont{\normalsize}
%
%



\vspace{2cm}
\begin{center}
{\large\textbf{Would Superluminal Influences Violate the Principle
of Relativity?}}\footnote{Forthcoming in \textsl{Lato
Sensu}, revue de la Soci{\'e}t{\'e} de philosophie des sciences.  This revised and corrected draft
was posted 19 May, 2013.  
} \\
\medskip
Kent A. Peacock\\
Department of Philosophy, University of Lethbridge.\footnote{4401
University Drive, Lethbridge, Alberta, Canada.  T1K 3M4. Email:
\url{kent.peacock@uleth.ca}}
\end{center}

\bigskip\bigskip


\begin{abstract}It continues to be alleged that superluminal
influences of any sort would be inconsistent with special relativity
for the following three reasons:   (i) they would imply the
existence of a `distinguished' frame; (ii) they would allow the
detection of absolute motion; and (iii) they would violate the
relativity of simultaneity.  This paper shows that the first two
objections rest upon very elementary misunderstandings of Minkowski
geometry and lingering Newtonian intuitions about instantaneity. The
third objection has a basis, but rather than invalidating the notion
of faster-than-light influences it points the way to more general
conceptions of simultaneity that could allow for quantum nonlocality
in a natural way.
\end{abstract}  

\bigskip




\tableofcontents

\newpage

\section{Alleged troubles with superluminal effects and influences}
There are at least two good reasons to take seriously the
possibility of superluminal influences.  First, they are arguably
though controversially implicated in the violations of locality
found in quantum mechanics.\footnote{\srr See, e.~g., Maudlin
\cite{Maudlin02} for a defence of this view.}
Second, it is not at all clear whether special relativity as it is
usually formulated actually \emph{excludes} superluminal influences
or simply \emph{fails to describe them properly}.\footnote{\srr In
support of the latter possibility I can only cite the large but
admittedly inconclusive body of literature exploring the possibility
of tachyons, superluminal frames, and extended relativity. It is not
possible to do a comprehensive review of this literature here. The
modern era of tachyon theory began in the 1960s, and papers from
that era by G.\ Feinberg \cite{Feinberg67} and O.-M.\ Bilaniuk and
E. C. G.\ Sudarshan \cite{BilSud69} are widely cited; see also
\cite{BDS62}. E.\ Recami insightfully advocated the significance of
tachyons in several publications; see his \cite{Recami86} for
comprehensive review. A neglected paper of 1986 by R.\ I.\
Sutherland and J.\ R.\ Shepanski \cite{SS86} is an important
milestone: their approach is different than the orthodox treatment
of tachyons by most authors. The orthodox approach is to substitute
the condition $v > c$ into the usual Lorentz transformations; this
gives an imaginary Lorentz factor $1/\sqrt{1 - v^2/c^2}$ leading to
many difficulties of interpretation. Sutherland and Shepanski show
that by re-deriving Lorentz-like transformations for the
superluminal case (rather than merely substituting into the
subluminal transformations) one arrives at a Lorentz factor
$1/\sqrt{v^2/c^2 - 1}$ which makes all proper quantities real-valued
for superluminal frames. M.\ Fayngold's recent review monograph
\cite{Fayngold02} on superluminal physics is very useful, but it
does not take account of Sutherland and Shepanski's important
innovation.  Very recently, J.\ M.\ Hill and B.\ J.\ Cox \cite{HC12}
have developed an `extended' version of special relativity that uses
the same real-valued Lorentz factor as Sutherland and Shepanski's
for $v > c$. }
To some knowledgeable observers these two claims will seem obvious,
to others they will seem to be `not even wrong.' I will not attempt
a detailed explication or advocacy of them in this paper, although
it will become clear where my sympathies lie. The major purpose of
this note is two-fold. First, it will attend to a job of undergrowth
clearance, which is to defend the notion of superluminality against
certain too-common misconceptions which stem from misunderstandings
of Minkowski geometry and lingering Newtonian intuitions about
simultaneity. Second, it will sketch a notion of simultaneity which
(properly developed) could allow for quantum-mechanical superluminal
influences in a natural way. This paper is propaedeutic to a larger
project being undertaken by this author that is aimed at defining
generalized conceptions of simultaneity which would be adequate to
the fact that we live in a quantum universe.

Some of the misunderstandings I will criticize here were dealt with
rather clearly by Frank Arntzenius \cite{Arnt90} over twenty years
ago. However, they continue to appear in the professional
literature, and so it seems necessary to respond to them yet again.
Let's begin with Barry Dainton, since the particular problems he
cites will be useful talking-points for my discussion. Dainton, in
his \textsl{Time and Space} \cite[p.\ 339]{Dainton10}, begins by
quoting J.\ R.\ Lucas \cite[p.\ 9--10]{Lucas90}, who said,
\begin{quote}
if some superluminal velocity of transmission of causal influence
were discovered, we should be able to distinguish frames of
reference, and say which were at rest absolutely and which were
moving.
\end{quote}
Dainton (p.\ 339) agrees, saying,
\begin{quote}
[a]nd in this he [Lucas] is surely right.  Were we to discover that
a truly instantaneous connection exists between objects at different
places in space, then assuming the connection has some detectable
effects, not only would the notion of absolute simultaneity have a
real application, but we would have a way of determining which
frames of reference are at absolute rest and which are not: \emph{it
is only with respect to frames truly} [sic] \emph{at rest that the
relative changes would occur at precisely the same time}. [Emphasis
added.]
\end{quote}

Ad{\'a}n Cabello, a distinguished researcher in quantum information
theory, has worries about instantaneity similar to Dainton's.  In
\textsl{Nature} Cabello reviewed recent findings on
quantum-mechanical correlations, and expressed his objection to
instantaneous influences in quantum mechanics as follows:
\begin{quote}
\dots the decision of what test is performed in one location cannot
influence the outcome of the test performed in the other location,
unless there is an instantaneous influence of the two tests on each
other.  \dots  But this is too high a price to pay, \emph{because it
is impossible to fit instantaneous influences into any theory in
which such influences travel at a finite speed} [emphasis added]
\cite[p.\ 456]{Cabello11}.
\end{quote}

Views like these have been held by other notable authors. Wesley
Salmon, for instance, argued that
\begin{quote}
[a]rbitrarily fast signals yield absolute simultaneity of the
strongest sort; the presence of the relativity of simultaneity in
special relativity hinges crucially upon the existence of a finite
upper speed limit on the propagation of causal processes and
signals \cite[p.\ 122]{Salmon80}.
\end{quote}
Nicholas Maxwell \cite[p.\ 38]{Maxwell85} seems to suggest that the
\emph{mere existence} of a superluminal effect would conflict with
the Principle of Relativity.  He cites his own interpretation of
quantum mechanics, which postulates the superluminal collapse of
spatially extended `propensitons' which, he argues, would explain
quantum correlations.
Maxwell insists that his own theory
\begin{quote}
irreparably contradicts special relativity.  For special relativity
asserts that all inertial reference frames are physically
equivalent.  In only one reference frame, however, will any given
probabilistic collapse of propensiton state be instantaneous; in
other, relatively moving frames the collapse will not, according to
special relativity, be instantaneous (though always
faster-than-light).
\end{quote}

From these remarks we can tease out four closely-related charges
against superluminality.  Before stating them, though, it will be
helpful to settle on terminology:
\begin{itemize}
\item A superluminal \emph{effect} will be any physical process that
    involves the faster-than-light propagation of a geometric locus such as the intersection
    point between a beam of light and a background,
    without presuming that this involves the transmission of any sort of
    influence or information faster than light.  A widely-discussed
    example is the searchlight-beam effect \cite{Rothman60,Weinstein06}:
    a beam of light from a rotating point source will track across a
    distant screen faster than light if the screen is at a sufficient
    distance.   Another example of a superluminal effect would be a
    string of   flashbulbs or firecrackers set to go off simultaneously
    in a   given inertial frame.  In all other inertial frames the
    sequence of   flashes or detonations will propagate superluminally;
    we'll return to the spacetime   kinematics of such processes shortly.
    It is usually taken that there is no question of   the searchlight
    beam effect or strings of firecrackers   transmitting
    \emph{influences} superluminally along their
    trajectories but it should not be assumed that the
    searchlight beam effect is unproblematic.\footnote{\srr  Rothman
    \cite{Rothman60} dismisses the possibility that the searchlight
    beam effect could transmit information along the trajectory of
    the intersection point, but Weinstein's wording \cite{Weinstein06} is more cautious,
    suggesting that the problem needs more investigation.
    H.\ Ardavan \cite{Ardavan84a} considered a scenario in which a rotating beam of
    electromagnetic radiation (such as that from a pulsar) traces a
    superluminal locus over a conductive surface (such as a layer of
    plasma spread through the solar system).  The beam of radiation will
    cause charge separation in the plasma and thereby induce
    electromagnetic radiation from the surface at the intersection locus.
    Ardavan carried out a rigorous calculation showing that if the locus
    orbits superluminally in a circular pattern then the induced field
    will diverge.  Ardavan further showed \cite{Ardavan84b} that the
    gravitational field induced by the searchlight beam effect in such
    scenarios will also diverge. Ardavan left it open whether these
    results indicate the high-field breakdown of classical
    electromagnetic and gravitational field theory, or some sort of
    otherwise-implausible prohibition on the searchlight beam effect.
    The questions raised by Ardavan's work remain importantly open.}
\item A superluminal \emph{influence} would be a hypothetical
  superluminal process in which some sort of causation passes faster
  than light from one point to another distant point.  I'd like to
  leave it as open as possible how superluminal influences, if any, would be
  constituted.
\item A \emph{tachyon} is a hypothetical faster-than-light particle,
  where we think of a particle as an entity that can be localized,
  at least under some circumstances.  I'll take a tachyon to be a form of
  superluminal influence, and I will sometimes use these terms
  interchangeably.\footnote{\srr R.\ Sigal and A.\ Shamaly \cite[p.\ 2358]{Sigal74} state, `we use the term
  tachyon to describe any propagation outside the light cone,' and
  this could include both of what I have called superluminal effects and influences.
  Sigal and   Shamaly's usage is not unusual in the literature; however,
  I will in this paper follow my narrower usage of `tachyon' because
  the three-fold distinction I sketch between different types of superluminal
  propagations allows me to make certain claims with less risk
  of misunderstanding.
  }
\item I will occasionally refer to superluminal influences, superluminal effects,
  and tachyons collectively as forms of \emph{superluminal propagation} when the difference between them
  doesn't matter.
\item Some papers in this literature (e.g. \cite{SS86} and references therein)
  speak of \emph{superluminal reference frames},
  which would be hypothetical faster-than-light Lorentzian physical
  systems which could be transformed \emph{to} in some versions of superluminal kinematics.
\item I'll also prefer the adjective `invariant' (`same in all frames of
  reference') to `absolute' or `truly' because that is more in keeping
  with standard usage in current relativity literature, and because it
  avoids dubious philosophical connotations.
\end{itemize}

Here are the Troubles with Superluminality with which we shall be
concerned:

\begin{description}

\item TS1:  There is no way to reconcile instantaneous influences
with `any theory in which such influences travel at a finite speed'
(Cabello).

\item TS2:  The existence of a superluminal physical influence (such as
Maxwell's propensiton collapse) would imply the existence of a
distinguished frame of reference and thereby violate the Principle
of Relativity (Lucas, Dainton, Maxwell).

\item TS3:  If a superluminal influence could be used to
transmit information controllably then it would be possible to
detect absolute states of motion and again thereby violate the
Principle of Relativity (Lucas, Dainton).

\item TS4:  If a superluminal influence were detectable or could be used to transmit
information controllably, it would allow violations of the
relativity of simultaneity (Dainton, Salmon).

\end{description}

I will show that TS1--3 rest upon elementary but surprisingly
widespread misconceptions about how superluminal motion would be
represented in Minkowski geometry.  As to the fourth (and much more
interesting) problem, I will have to respond \emph{guilty as
charged}, but I will argue (though not as conclusively as with
TS1--3) that the charge is not nearly as damaging as most people
suppose, and that it may in fact open a door to interesting new
physics.

\section{Superluminal propagation does not imply a distinguished frame}
\noindent Before we address the implications of the detectability of
superluminal influences, let's review some basics of superluminal kinematics
in special relativity. Consider a familiar spacetime diagram, restricted to
2-dimensional $(x,t)$ space for simplicity:


\begin{center}
\includegraphics[scale=.4]{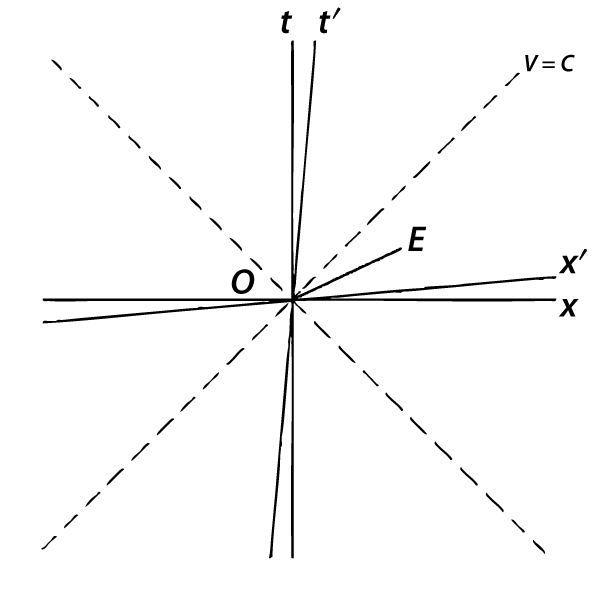}\\
Fig.\ 1
\end{center}

We'll take the $x$  and $t$  axes to be the space and time axes
respectively of the `lab' frame $S$ and the $x'$  and $t'$ axes to
be the space and time axes of another inertial frame $S'$ moving
subluminally to the right. We've chosen units such that the light
cone is at $45^\circ$  with respect to the $x$ and $t$ axes, and the
line $OE$ is the trajectory of something propagating with constant
superluminal velocity to the right. Geometrically, what defines any
form of superluminal propagation is that its spacetime trajectory is
outside the light cone as shown. The line $OE$ therefore need not be
the worldline of an exotic hypothetical particle, a collapsing
propensiton, or the Starship \textsl{Enterprise} moving at warp
speed; it could simply be a string of flashbulbs timed to go off
simultaneously in some inertial frame. If the relative velocity of
the moving $(x',t')$ frame steadily increases, the $x'$ and
$t^\prime$ axes rotate uniformly toward the light cone in order to
preserve the invariance of the speed of light for both frames.  (Of
course, the proprietors of the $(x^\prime, t^\prime)$   frame are
perfectly entitled to draw \emph{their} axes as orthogonal and the axes of
the lab frame as rotating away from the light cone.) Recall that the
spatial hyperplanes of a Lorentz frame serve as its hyperplanes of
simultaneity when simultaneity is defined according to Einstein's
clock synchronization convention (see Taylor and Wheeler
\cite{WT66}): any string of events along a line parallel to the
$x$-axis of $S$ is simultaneous in $S$ (though not in any frame
moving with respect to $S$). As the $x^\prime$-axis rotates toward
the light cone, there will be exactly one relative velocity between
lab and moving frames at which the spatial axis of the moving frame
coincides with $OE$, and just as Maxwell says, in this frame \emph{and
this frame only} the propagation along $OE$ will be instantaneous
(though it is superluminal in all frames). Indeed, if $u$ is the
velocity of a superluminal propagation with respect to the lab
frame, then this propagation moves infinitely fast not in a frame
`truly at rest' as Dainton has it, but in a frame moving with
velocity $v=c^2/u$ with respect to the lab frame.\footnote{\srr This
follows from Lorentz transformation for time: $\D t' = 0 = (\D t -
u\D x/c^2)$ implies $\D x/\D t = v = c^2/u$.  See Rindler
\cite{Rindler79}, especially pp.\ 90--91. }

There are thus two salient facts about instantaneity in special
relativity:  first, instantaneity with respect to a frame of
reference is a perfectly admissible concept; second, there is no
invariant concept of instantaneity if `instantaneous' means
`traversing a distance with no lapse of coordinate time'. (The
frame-dependence of instantaneity is merely the relativity of
time-coordinate simultaneity in different words.)  There is no
question that the frame-dependence of infinite velocity clashes with
Newtonian intuitions that are hard to dislodge. However, because
instantaneity (or equivalently infinite velocity) is a
frame-dependent concept, there is no way that any form of
superluminal propagation (even though it is necessarily infinitely
fast in some frame, as shown in Fig.\ 1) could define an absolute
rest frame, a notion that cannot even be represented in the
mathematics of special relativity (let alone on a spacetime
diagram).

Another way to look at it is to note that any ordinary object moving at some
velocity less than the speed of light defines a special frame as well, namely
its local co-moving rest frame.  No one supposes that the fact that every
subluminal object is at rest in its own private inertial frame picks out a
`privileged' frame whose existence threatens the Principle of Relativity. The
local co-moving frame of a subluminal particle is determined by the
contingent details of that particle's history, and is not by itself a
universal law of physics. Similarly, no one need suppose that if there is so
much as one instance of superluminal propagation in the universe then its
existence would pose a threat to the Principle of Relativity just because its
motion is instantaneous in a particular frame of reference. Again---any such
frame of instantaneity would be picked out not as a matter of universal law
but as a consequence of the accidents of the dynamical history which led to
that particular propagation.

It is essential to grasp that while the Principle of Relativity
requires that there be a covariant description of every possible
physical process, it does not imply that everything looks the same
in every admissible state of motion.\footnote{\srr What I say here
does not add much to the very clear argument given by F.\ A.\ Muller
\cite{Muller92}. } As noted, any discrete object is at rest only in
its own local co-moving rest frame. Another pertinent example is the
electromagnetic field; it has a covariant description (see, e.~g.,
Misner, Thorne, and Wheeler \cite[\S 3.4]{MTW73}) but this surely
does not mean that any given electromagnetic field looks the same in
all states of motion.  For example, the field of a point charge in
its own rest frame has no non-zero magnetic components, but this
hardly implies that the rest frame of a charge is `privileged' (even
though when doing electromagnetic theory it is often useful to
simplify a problem by finding a frame, if one can, in which some
components of the field vanish). Similarly, \emph{pace} Maxwell, the
fact that any superluminal influence is infinitely fast in one but
only one frame does not contradict the Principle of Relativity.

There is a subtle fact about relative velocities that is not always
explicitly mentioned in books on relativity, and a failure to grasp
this subtle fact may be a cause of some of the confusion about
superluminal motion. In special relativity all velocities (except
for the velocity of light itself) are relative, \emph{including}
zero and infinite velocity. However, it is an \emph{invariant} fact
whether or not two physical systems have a certain \emph{relative}
velocity.  Thus it would be an invariant fact whether or not a
certain tachyon beam is instantaneous relative to a certain inertial
frame. Perhaps this is part of what has puzzled those who apparently
believe that the mere existence of superluminality would imply the
existence of an invariant or `absolute' state of motion other than
the motion of light itself.  Dainton \textit{et al.} possibly have
confused the \emph{invariant} fact that any superluminal propagation
has infinite velocity relative to one frame (which one depends on
the spacetime trajectory of the superluminal effect) with the notion
(not correct) that any superluminal propagation would be invariantly
infinite for \emph{all} frames.

When Dainton speaks of `truly' instantaneous connections his usage
is ambiguous.  No connection outside the light cone is instantaneous
in more than one physical frame although in that frame it is `truly'
instantaneous. Which frame it is depends upon initial conditions and
is not some law of nature. And since any instantaneous connection is
\emph{superluminal} in all frames the question of instantaneity is a
red herring; the real question is what we are to make of
superluminal influences.

Let's go back to Cabello's worries about Bell correlations. There is
no question that if they are due to any sort of causal influence it
must be superluminal, and if it is superluminal in one frame it is
superluminal in all. However, that hardly implies that such
superluminal influences would be instantaneous in \emph{any} given
frame. Which frames they happen to be instantaneous in would depend
upon the initial and boundary conditions of the experiment. Thus,
there certainly is a theory that allows for influences which are
instantaneous in one frame and finite (though superluminal) in all
others: it is called `special relativity.'

\section{Superluminal influences would not allow detection of absolute motion}
In order to address TS3, let's consider a slightly more complicated
scenario. In Fig.\ 2, Alice and Bob are localized observers moving
through spacetime. They were initially coincident at $O$ and at that
point they synchronized their local co-moving standard clocks.  As
the diagram suggests, they undergo varying accelerations in their
careers through spacetime.  If they are brought back into
coincidence at a much later point it will be found that in general
their elapsed proper times (given by the readings on their co-moving
clocks) will differ. This is the much-debated Twin Paradox, which is
based on the fact that elapsed proper time is path-dependent but
invariant while coordinate time (as defined by Einstein's clock
synchronization convention) is global but frame-dependent. (See
Marder \cite{Marder71}, Arthur \cite{Ric08}, and H.\ R.\ Brown's
lucid discussion of clocks as the `waywisers' of spacetime \cite[p.\
95]{HRB06}.)
%
\begin{center}
\includegraphics[scale=.35]{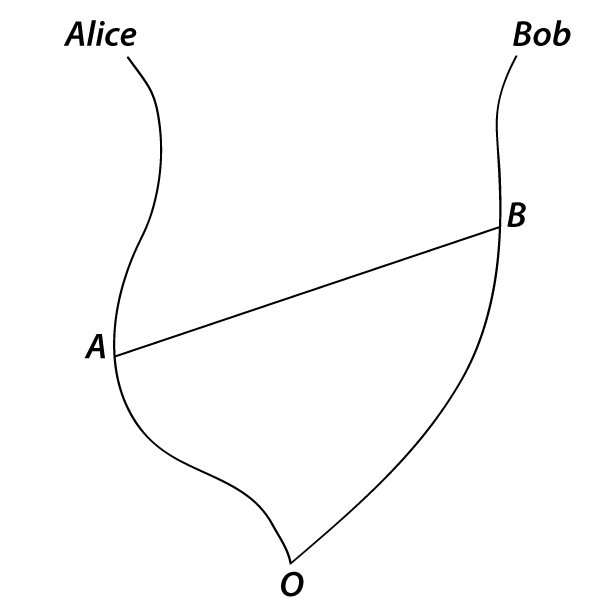} \\
Fig.\ 2
\end{center}

Suppose that Alice emits a tachyon beam at $A$ which moves with
constant superluminal velocity until it happens to intersect Bob's
worldline at his world point $B$.  I say `happens to' since I'm not
appealing here to any speculative theory that would give rules
governing the motion of tachyons; for all we know, Alice's tachyon
beam could have been emitted in a random direction in spacetime and
the fact that it intersects Bob's worldline at $B$ could be pure
chance. Nevertheless, it is invariant that the beam intersects
Alice's worldline at $A$ and Bob's at $B$ and that it follows a
certain trajectory through spacetime between these world points. The
ordering of $A$ and $B$ with respect to a global time coordinate is
frame-dependent, but the fact that these points are connected by the
tachyon beam is not.

If Alice and Bob happen to be (even momentarily) at rest with
respect to each other at the points $A$ and $B$, then they share a
common inertial frame, which can be defined so that the line $AB$ is
its spatial $x$-axis.  The tachyon connecting $A$ and $B$ will be
instantaneous in this frame (and, strictly speaking, in any frame
related to it by mere translation). This, at the risk of repetition,
is certainly an invariant fact. But if Alice and Bob are linked by a
tachyon beam which is instantaneous in their mutual rest frame, that
does not in the slightest degree imply that Alice and Bob are
`absolutely' at rest, as Lucas and Dainton seem to think. The
invariance of a state of relative rest does not imply the existence
of a globally invariant state of rest any more than does the
invariance of the fact that Alice and Bob could be moving at some
non-zero finite velocity  $v$ with respect to each other imply that
$v$  is an absolute velocity in any sense that would have interested
Newton.  Again, the mere fact that Alice and Bob might be connected
by a tachyon beam---or for that matter a string of flashbulbs which
happen to have been arranged so as to pop off simultaneously in
Alice and Bob's mutual rest frame because \emph{kinematically} these
things are equivalent---surely does not by itself imply that they
are at rest in any `absolute' sense. This is despite the fact that
it could be an invariant fact that they are \emph{relatively} at
rest.

Now, what about detectability?  Let us imagine what some might say
would be the worst case scenario, which would be that Alice can
`ping' Bob by means of a readable tachyon signal along $AB$ and Bob
can bounce a readable response back to Alice along $BA$ with no
lapse of proper time for her\footnote{\srr I have not said anything
about the elapsed proper time for the \emph{tachyon} between $A$ and
$B$.  On the orthodox reading of special relativity, proper time for
spacelike propagations is imaginary since $ds^2 < 0$ outside the
light cone.  However, in the superluminal kinematics developed by
Sutherland and Shepanski \cite{SS86} $ds^2 \geq 0$ for all
intervals, spacelike and otherwise. It is beyond the scope of this
paper to adjudicate between their view and the orthodox view.
Suffice it to say that how we parameterize proper time along the
tachyon's path is not relevant to our discussion here. }
between transmission and reception, and with a complete picture of Bob's
local physical state at $B$ encoded in the return signal to Alice. The
tachyon probe would make it as if Alice at $A$ could be momentarily
coincident with Bob at $B$. Alice can therefore learn exactly as much but no
more from the tachyon signal than she could if she and Bob's worldlines
happen to cross at $A$ and $B$. Even with this much information about Bob,
the very most that Alice could know about Bob's velocity at $B$ is his
\emph{relative} velocity with respect to her at $A$. Why? Because that is all
the information about Bob's state of motion there is to be had---Bob doesn't
\emph{have} an absolute velocity. Alice can use the tachyon beam to determine
the \emph{invariant} fact of her \emph{relative} velocity with respect to
Bob, but if she understands relativity theory she will not be confused by the
fact that it is invariant whether she and Bob are relatively at rest at
certain points.

Indeed, Alice need not have used tachyon beams at all to know her
state of motion with respect to Bob's at $A$ and $B$, for she and
Bob (when they were coincident at $O$) could have arranged in
advance that they would follow acceleration schedules such that they
would be relatively at rest at points $A$ and $B$.
No one would dream of suggesting that the fact that they could do
this would define an absolute state of rest that would violate the
Principle of Relativity.  There is no good reason at all to suppose
that the mere fact that Alice and Bob can somehow infer that they
are mutually at rest at some spacetime points or other implies the
existence of an absolute or invariant state of rest, and this fact
is independent of whether they are connected by a tachyon that
happens to be instantaneous in their mutual rest frame, whether the
`connection has some detectable effects,' or whether Alice at $A$
and Bob at $B$ have any way at all of measuring directly or
inferring each other's states of motion.

There's another way of looking at it.  Suppose the point $O$ is at
rest in the `lab' frame, and suppose as before that Alice and Bob
are momentarily at rest with respect to each other at points $A$ and
$B$. At those points they could be moving at any velocity from zero
to arbitrarily close to but not equal to $c$  with respect to the
lab frame.  There are therefore indefinitely many velocities with
respect to the lab frame at which Alice and Bob could be at rest
with respect to each other.  Hence the fact that they can be at rest
with respect to each other can hardly define a unique state of rest,
which would surely have to be unique if it were indeed `absolute.'
Again, this is completely independent of whether or not Alice or Bob
could use tachyons or any other means to tell that they were
relatively at rest at $A$ and $B$.

These observations help to explicate a remark made by Arntzenius
\cite[pp.\ 229--230]{Arnt90}):
\begin{quote}
  When W.\ Salmon [claims] that tachyons, if they could be used as
  signals, could establish absolute simultaneity, he does not indicate
  how one could do this.  Assuming the frame-independence of the speed
  of light, and the frame dependence of the speed of tachyons this in
  fact appears to be a hopeless project:  which tachyons exactly are
  to be used to establish absolute simultaneity?
\end{quote}
What I am mostly doing here in my response to TS3 is to spell out in
almost painful detail this point made by Arntzenius in
1990.\footnote{\srr Arntzenius' paper `Causal Paradoxes in Special
Relativity' \cite{Arnt90} is still an essential prerequisite for
anyone who wishes to investigate the puzzles arising from the
possibility of causal looping in special relativity.  Where we can
go beyond Arntzenius today is that more is known about entanglement
and extended versions of quantum mechanics such as the two-state
formalism; these topics, which open the door to decidedly
non-classical notions of causation, are discussed briefly in later
sections of the present paper. }
 With apologies to Arntzenius, it seems that this
point needs to be made again, with as much clarity as can be
mustered. Again, if I may: \emph{Any} tachyon is instantaneous with
respect to \emph{some} frame and there is no basis on which to pick
one tachyon as definitive of an invariant state of rest. Conversely,
for \emph{any} inertial frame there is a trajectory which is
instantaneous in that frame; which of the indefinitely many such frames do we pick as
privileged?

\section{Superluminal influences would not conflict with Einstein's relativity of simultaneity}
It should be clear that TS1--3 can be obviated by a bit of careful thought
about how spacetime diagrams work. But now we must say something about
Dainton's worry (TS4) about distant clock synchronization, which raises much
more interesting difficulties---and possibilities. Precisely what can Alice
and Bob do with tachyons that they cannot do with light signals?

First, if a readable signal, \textit{per impossibile} perhaps, could be
imposed on a tachyon beam, then Alice and Bob could momentarily synchronize
their local clocks at the points $A$ and $B$. What I mean is that if Alice
can send her local clock reading at point $A$ to Bob at point $B$ then Bob
could set his local clock reading at $B$ to agree with Alice's local reading
at $A$. Our Newtonian intuitions prompt us to think that the distant clocks
are synchronized only if they have the same reading \emph{at the same global
time coordinate}.  However, this has no invariant meaning in special
relativity, whereas it is invariant whether the local readings at $A$ and $B$
are equalized as described.  Whether or not the local clock readings are
equal at $A$ and $B$ is therefore independent of whether or not $A$ and $B$
are at the same global time coordinate in some inertial frame or other.

Distant clocks in special relativity can be synchronized using light
signals but it takes a certain minimum amount of time to do that in
\emph{every} frame.  The new thing that Bob and Alice could do with
controllable tachyons is synchronize their distant clocks so that
there exists an inertial frame in which the process takes \emph{no}
time.  It could be done by picking the frame in which $AB$ is a
spatial axis connecting Bob and Alice. The events $A$ and $B$ are at
the same time \emph{in this frame} and the tachyon signal will be
instantaneous \emph{in this frame}.  But whether or not the clocks
at $A$ and $B$ are synchronized by tachyons interchanged between $A$
and $B$ is completely unaffected by any choice of inertial frame in
which the process is described.

I said that Alice and Bob could synchronize their clocks
`momentarily' because unless they happen to remain at rest with
respect to each other their local clocks would get out of synchrony
again as they continue their careers through spacetime. On the other
hand, if Alice and Bob continue to stay at rest with respect to each
other then their clocks, once synchronized by tachyons at $A$ and
$B$, would stay in synchrony.

The crucial point of this story is that Alice and Bob's ability to
synchronize their clocks using tachyons would not violate the
relativity of simultaneity defined as equality of a global time
coordinate, because the latter is based upon Einstein's
considerations about how one could synchronize clocks using light
rays given that the speed of light is both finite and invariant.
Einstein's way of defining time-coordinate simultaneity neither
assumes nor requires that light signals be either the fastest or the
only way of communicating between distant events; it's only about
what can be accomplished with light signals.\footnote{\srr
The persistent belief that relativity is based upon the assumption that
light is a `first signal' \cite{Reich57} is arguably an instance of
what John Woods \cite[p.\ 153]{Woods03} has described as the
\emph{Heuristic Fallacy}:
\begin{quote}
  Let $H$ be a body of heuristics with respect to the construction of
  some theory $T$.  Then if $P$ is a belief from $H$, which is
  indispensable to the construction of $T$, then the inference that
  $T$ is incomplete unless it sanctions the derivation of $P$ is a
  fallacy.
\end{quote}
(Woods goes on (p.\ 154)  to explain diplomatically that many
fallacies are errors `that even the attentive and intelligent are
routinely disposed to make.') While the notion that $c$ is a
limiting velocity certainly played a key historical role in
motivating the construction of special relativity by Einstein and
Poincar{\'e}, this notion is not formally used as a premise of the
theory, and it is only debatably a theorem of special relativity.}
Alice and Bob could have tachyon-based radios and yet still go ahead and
set up a coordinate system using ordinary laser beams and Einstein's
synchronization procedures (as in, e.~g., Wheeler and Taylor
\cite{WT66}), and all the strictures identified by Einstein would
continue to apply to the latter. The possibility of tachyon signals
makes no difference to the relativity of time-coordinate
simultaneity, for it simply would give \emph{another way} of
coordinating distant events than by means of electromagnetic
signals.  This point was made by G.\ Nerlich quite some time ago
\cite{Nerlich82} but it seems that it must be made again.

Thus Salmon's statement that `the relativity of simultaneity \dots
hinges crucially upon the existence of a finite upper speed limit on
the propagation of causal processes and signals' \cite[p.\
122]{Salmon80} is simply incorrect.  The relativity of
time-coordinate simultaneity (where times are defined using the
synchronization procedure recommended by Einstein in 1905
\cite{Einstein05}) and indeed the entire mathematical structure of
special relativity is dependent upon the assumption that the vacuum
speed of light is a finite \emph{invariant}, not necessarily a
maximum. It seems clear that Einstein himself believed that $c$ is a
universal speed limit, and it is also clear that many authors would
\emph{prefer} that this were the case,\footnote{\srr
An important example is J.\ S.\ Bell, who said that by his theorem
`maybe there must be something happening faster than light,
\emph{although it pains me even to say that much}' [emphasis added];
\cite[p.\ 90]{MC88}.}
but that assumption is not mathematically required in order to
derive the Lorentz transformations.  To confirm this, review
Einstein's own derivation of the transformations \cite{Einstein05},
or see any standard presentation of special relativity (e.g.,
Wheeler and Taylor \cite{WT66}).

If there are, indeed, superluminal influences or connections of some
sort, then there is no good reason to think that they could not
peacefully coexist in parallel with Einstein's time-coordinate
simultaneity.\footnote{\srr
In this paper I have entirely skirted the large and subtle
literature on the conventionality of simultaneity, because that
problem is about what can be accomplished with light signals---an
important question that is orthogonal to my interest here, which is
to explore what could be accomplished with other sorts of signals
than light. For an up-to-date review of the conventionality of
simultaneity, see \cite[pp.\ 95--105]{HRB06}.)}

\section{But superluminal influences might allow alternative concepts of
simultaneity}
What would superluminal influences, controllable or otherwise, add
to our understanding of simultaneity?  Is there any sense in
speaking of superluminal influences as definitive of
simultaneity-like relations on spacelike-separate events?

The problem is that even though everyone knows that simultaneity
defined in terms of global time coordinate is frame-dependent,
almost everyone still wants time-coordinate simultaneity to do the
same metaphysical work that absolute-time simultaneity does in
Newton's universe.  Newton's absolute time is the great steady
heartbeat of his universe, and all physical changes in that universe
are with respect to it.  In Einstein's universe there are
indefinitely many ways of coordinatizing events; none are
metaphysically privileged though some may be preferable for
practical reasons. Einstein's procedures for setting up space and
time coordinates using light signals and standard measuring rods is,
to be sure, very useful (in large part because it nicely reduces to
the Newtonian picture in the limit of low relative velocities), but
it is only one possible way of painting coordinates onto events;
general covariance tells us that no coordinatization of events is
privileged in any physical or metaphysical sense
\cite{Rovelli04}.\footnote{\srr In many relativistic cosmologies,
such as Robertson-Walker universes, there can be a global time, but
it is \emph{history-dependent} and does not conflict with general
covariance. See, e.g., \cite{Weinberg08}. }
It is therefore not automatically given that any physical
connectivity or equivalence between spacelike separate events must
be described in reference to hyperplanes of constant time
coordinate.

This, by the way, is the blind spot that has dogged discussions of
the problem of finding a covariant description of quantum state
reduction. Even the best-informed authors in this literature (e.~g.,
Aharonov and Albert \cite{AA80,AA81}) assume that wave function
collapse has to occur over hypersurfaces of constant time
coordinate, which leads to the immediate conclusion that there is no
covariant description of the process, if it is a physical process at
all.\footnote{\srr  According to Aharonov and Albert,
\begin{quote}
[i]n the nonrelativistic case a measurement is taken to set initial
conditions for the propagator over the equal-time hypersurface of the
measurement event\dots\ In the relativistic case, however, different
observers will in general have different definitions of this
hypersurface\dots\ different observers may derive different sets of
probabilities.  \cite[p.\ 3322]{AA80}
\end{quote}
They go on to explain that there is, after all, a consistent way of
predicting the probabilities of local measurement results, with the
aid of microcausality, but `a description of the physical system in
terms of its observables simply cannot consistently be written down'
\cite[p.\ 3324]{AA80}.  But if state reduction is superluminal,
which it must be, then for the elementary kinematic reasons
explained in this paper there is only one frame in which it could
reset probabilities over an equal-time hypersurface. Therefore, it
is just a mistake to suppose that every observer would describe the
reduction process as instantaneous.}
If one were to seek a covariant description of state reduction, one
would want to see if this can be done in terms of covariant
properties of the wave function.\footnote{\srr  It is important to
say what sort of wave function we are discussing when we talk of the
problem of finding a covariant description of wave function
collapse.  The wave function is not \emph{necessarily} an object
living in configuration space.  A wave function in general is simply
the projection of the state function into a continuous
representative \cite[Ch.\ II, \S E]{CT77vI}, which could be
configuration space, ordinary spacetime, or momentum-energy space.
What I am talking about here, and what most of this literature
concerns itself with, is the de Broglie wave packet, which is a
projection of the state function into Minkowski space; see Dirac
\cite[\S 30]{Dirac58} for a succinct review of the de Broglie wave.
}
An obvious candidate is \emph{phase}:  it is far more natural to
think of wave functions as reducing over hypersurfaces of constant
phase, and this automatically gives a covariant picture; given
appropriate initial conditions, it may also be possible to describe
state reduction in terms of constant \emph{action}
\cite{Riet85,Peacock06}. These proposals require much technical
development but, from the spacetime point of view advocated in this
paper, the conventional assumption that state reduction is linked to
hypersurfaces of constant time coordinate seems to be among the
\emph{least} promising approaches to the problem.

Returning to the problem of simultaneity, consider Fig.\ 2, and
again suppose there is some sort of connection or influence outside
the light cone between points $A$ and $B$. This would most likely be
quantum mechanical in its basis, but to see the point I want to make
we need not worry about the precise nature or origin of this
connection; whether or not there are such influences or connections
is an empirical question which cannot be settled on an \emph{a
priori} basis from the postulates of special relativity as they
presently stand.  Whatever the details of the dynamics may be, the
\emph{kinematics} of such connections is clear in the following
respect:  the connection between $A$ and $B$ is \emph{factual} in
the sense that all observers in all states of motion will agree that
it is \emph{those} two points, $A$ and $B$, which are connected in
this particular way; the fact that these points are connected is
relativistically invariant.  As we have seen, $A$ and $B$ will be at
the same time in one and only one frame, which (again) is
`distinguished' only by its dynamical history and not by some law of
nature. In all other frames $A$ and $B$ will be at different time
coordinates, and so whether or not two spacetime events are
connected in this peculiar invariant but history-dependent way has
nothing to do with whether or not they are at the same time in some
Lorentz frame.

I would now like to suggest that if such connections do exist it is
meaningful to say that they define a kind of simultaneity relation between
$A$ and $B$---though obviously not the sort of simultaneity defined by
Einstein, which is based on equality of a global time coordinate.  A full
treatment of this question is beyond the scope of this paper, but I'll try to
say enough to show where this inquiry could go.

The key is that the modern usage of the term `simultaneity' equivocates on
two distinct senses of the term.  According to Max Jammer \cite{Jammer06},
the etymological root of `simultaneity'
\begin{quote}
is, of course, the Latin ``simul,'' which in turn derives from the Sanskrit
``sem'' (or ``sema''), meaning ``together,'' both in the sense ``together in
space'' and ``together in time'' \cite[p.\ 11]{Jammer06}.
\end{quote}
The \textsl{Oxford Latin Dictionary} \cite{OLD} tells us that the
Latin \textit{simul} has two distinct senses: two events may be
\textit{simul} if they occur at the same time, but events may also be
judged \textit{simul} if they are in some way \emph{together} or
\emph{in joint process}---that is, part of some larger or more
extensive coherent whole. Our events $A$ and $B$ are \textit{simul}
in the second sense in all frames of reference, but \textit{simul}
in the first sense in only one.  The notion of simultaneity as joint
process is an epistemically more more primitive sense of
simultaneity than simultaneity in terms of time coordinate, since
judgements of time are built up from judgements of coincidence
(localized joint process) between clock readings and localized
events. In a Newtonian universe it is natural to assume that events
in joint process are at the same absolute time, but this does not
follow in an Einsteinian universe.\footnote{\srr
A small number of authors have explored the notion that there are distinct
senses of simultaneity.   Adolph Gr{\"u}nbaum \cite[p. 203]{Grun73} defined
what he called \emph{topological} simultaneity:  events simultaneous in this
sense are those that cannot be connected causally.  Since he thought that any
sort of spacelike causal connections are excluded by relativity theory, all
events spacelike separate from $A$ are topologically simultaneous with
respect to it. Whether or not events are topologically simultaneous is an
invariant distinction. Brent Mundy \cite{Mundy86} similarly defined what he
called \emph{causal} simultaneity as the absence of any possible causal
connection, but unlike Gr{\"u}nbaum he argued that relativity does not
logically exclude the possibility of spacelike causal connections; therefore,
on Mundy's view, the sets of causally simultaneous events might not comprise
the whole region outside the light cone.  The synchronization of distant
clocks according to Einstein's clock synchronization convention was called by
Mundy \emph{optical} simultaneity. Mundy argued that the presentations of
relativity by Gr{\"u}nbaum and Reichenbach \cite{Reich57} are
reconstructions, based on the unnecessarily strong assumption that light is a
`first signal,' which distort the meaning of the theory and drastically limit
its scope.}

In orthodox relativity the notion of invariant joint process is
accepted so long as the events are coincident.  In Einstein's words,
\begin{quote}
We assume the possibility of verifying `simultaneity' for events
immediately proximate in space, or---to speak more precisely---for
immediate proximity or coincidence in space-time, without giving a
definition of this fundamental concept \cite[p.\ 115]{Einstein16}.
\end{quote}
But even in his earliest writings on the theory of relativity
Einstein was well aware that the notion of the coincidence of two
presumably point-like events is neither mathematically nor
physically clear:
\begin{quote}
We shall not here discuss the inexactitude which lurks in the
concept of simultaneity of two events at approximately the same
place, which can only be removed by an abstraction \cite[p.\
39]{Einstein05}.
\end{quote}
Up to now I have been largely concerned with pointing out the
respects in which superluminal influences are consistent with
relativistic kinematics as presently understood.  We now reach a
boundary beyond which one must consider ways in which relativity
needs to be expanded to take account of quantum mechanics.  Physics
can no longer avoid addressing the `lurking inexactitude' cited by
Einstein.

In classical relativity the ambiguity in the notion of infinitesimal
closeness is simply ignored; spacetime is taken to be built up out
of point-like events and if these are spacelike separate they are
presumed to be causally disjoint (except insofar as they can be
linked by backwards and forwards light cones). Therefore, from the
point of view of quantum mechanics the concept of an event in
classical relativity is ambiguous in two respects. First, the
physical meaning of coincidence or infinitesimal closeness is
unclear.  This is partially because of the Uncertainty Relations;
also, some current approaches to quantum gravity (e.g.,
\cite{ADV09}) open up the possibility that space and time may be
discrete at the Planck scale. If spacetime is discrete then even
events separated by one quantum of length are spacelike separate
and, by the classical criteria, could not be considered coincident.
Second, and most pertinent to the theme of this paper, is the vexing
question of whether events outside each other's light cones are
causally disjoint. No one doubts that any collection of events can
be associated by convention in an essentially arbitrary way; the
question is whether it makes any sense to speak of \emph{distant}
events as being in `joint process' in a causal or dynamical way that
is somehow demanded by the physics of the situation.

There is increasing evidence that quantum mechanics shows that
distant particles, especially if they are entangled, may be
nonseparable or form or partake in a unity in surprising ways.  It
may therefore be sensible to generalize the conception of an event
to allow for events and states that are extended throughout
spacetime in an invariant way.

A dramatic example of the inseparability of spatially extended
quantum states appears in a recent experiment by K.\ C.\ Lee \textit{et al.}\
\cite{LeeKC11}.  These experimenters used a complicated
interferometric apparatus in which two 3 mm diamond chips separated
by 30 cm were put into entangled phonon states (phonons are quanta
of vibrations) and then `pinged' by an ultra-high frequency laser.
The key point for our discussion here is that the diamond chips were
demonstrably put into a \emph{single} quantum state despite their
spatial separation.  As Lisa Grossman explains,
\begin{quote}
[t]o show that the diamonds were truly entangled, the researchers
hit them with a second laser pulse just 350 femtoseconds after the
first.  The second pulse picked up the energy the first pulse left
behind, and reached the detector as an extra-energetic photon.  If
the system were classical, the second photon should pick up extra
energy only half the time---only if it happened to hit the diamond
where the energy was deposited in the first place.  But in 200
trillion trials, the team found that the second photon picked up
extra energy every time.  That means that the energy was not
localized in one diamond or the other, but that they shared the same
vibrational state \cite{Grossman11}.
\end{quote}
It is as if quantum mechanics simply does not know or care that the
two diamond chips are 30 cm apart.  Lee \textit{et al.}\ do not
attempt a covariant description of their nonlocal energy states but
their result is an example of the sort of scenario we discuss here:
at certain proper times along their world-lines, the two diamond
chips share a certain common energy state.  Whatever the detailed
spacetime description may be (this remains to be worked out) it has
to be an invariant fact that \emph{those} points on their
world-lines are linked in \emph{that} particular invariant and
nonseparable manner.  Most important, the single nonlocal energy
state shared by the two distinct diamond chips is demonstrably not
reducible to two local energy states possessed by the two chips.
That's an important part of what it means to say that the state is
entangled: it is not separable into distinct and localizable
sub-states.  To be sure, the diamond chips have other physical
properties which are localizable in the normal way, but their
non-separable, spatially-extended energy state seems to be a very
natural candidate for an entity that is \emph{simul} in the second
sense. One cannot avoid speaking of it as being in `joint process'
because it cannot even be analyzed into distinct localized parts.

It is quite likely that such alternative notions of simultaneity as
suggested here---invariant but history-dependent---would violate
some people's intuitions about a vaguely-defined `Spirit of
Relativity,' but it is not obvious that they are not allowed by the
mathematical structure of relativity and they seem to be demanded by
quantum physics.  While relativity is far more amenable to
superluminal influences that has been generally supposed, ultimately
it is classical relativity that must adapt itself to the quantum
\cite{CC93,Peacock98}.

To summarize:  if we grant that there could be invariant connections
between spacelike separate events, likely quantum mechanical in
their basis, then it is reasonable to call it a kind of simultaneity
relation because it answers to the notion of distant events as being
part of a single process.  Quantum mechanics \textit{prima facie}
demands that we disambiguate the two key senses of simultaneity that
have been conflated since the time of Newton.

\section{Are `causal' accounts of quantum mechanics consistent with the Principle of Relativity?}
An anonymous referee for this paper made a very helpful observation:
\begin{quote}
  [T]here are theories that are phenomenologically compatible with
  special relativity  in which superluminal propagation does pick out
  a preferred frame.  Bohmian mechanics (also referred to as
  `pilot-wave' theory or de Broglie-Bohm theory) has a preferred frame
  of reference.  Perhaps theories like these are feeding the
  intuitions of those making claims akin to TS2\dots
\end{quote}
This is quite likely right.  For instance, Maudlin \cite{Maudlin02}
argues that Bell's Theorem could force us to concede that there is a
special frame which is preferred although \emph{undetectably} so.
Thus one must ask whether any theory that attempts to underpin
quantum statistics by means of nonlocal dynamics is
\emph{necessarily} in conflict with Lorentz invariance.  Or to turn
the question around, can there be a \emph{covariant} theory of
nonlocal dynamics?

Bohm's `hidden variable' theory of 1952 \cite{Bohm52,Cushing94} is
Galilean invariant because Bohm never intended it to be otherwise;
his aim was to show that non-relativistic wave mechanics could be
underpinned by a causal (though unavoidably superluminal) dynamics
in which particles apparently have definite trajectories.  Hence it
is reasonable to investigate whether a relativistic generalization
of Bohm's theory is possible.  Bohm himself apparently thought not:
he and Basil Hiley state that `it would be extremely surprising to
obtain a Lorentz invariant theory of particles that were connected
nonlocally' \cite[p.\ 282]{BH93}.  They consider two spacelike
separate particles $A$ and $B$, `both at rest in the laboratory
frame' at worldpoints $a$ and $b$ respectively, and then remark,
\begin{quote}
  [i]f there is a nonlocal connection of the kind implied by our
  guidance condition, then it follows that, for example, points $a$ and $b$
  instantaneously affect each
  other.  But if the theory is covariant, there should be similar
  instantaneous connections in every Lorentz frame.
\end{quote}
Their accompanying figure shows connections from $a$ to other points
on $B$'s worldline. Although their language is unclear, Bohm and
Hiley do seem to grasp that each possible spacelike connection
between $a$ and the points along the worldline of $B$ would be
instantaneous in one and only one Lorentz frame; there is no
covariant sense in which \emph{all} are instantaneous.  However,
they go on to say that from the fact that there could be
instantaneous connections between $a$ and earlier points on $A$'s
own worldline via points on $B$'s worldline, it would be possible to
set up a typical closed-loop causal paradox in which an influence
from $a$ could interfere with $A$'s own history at an earlier
worldpoint along  $A$'s worldline in such a way as to prevent the
influence from being emitted at $a$.

Closed causal loops are a genuine problem for superluminal theories,
but the risk of a closed causal loop has nothing to do with whether
or not the connections are instantaneous in some frame or other, for
that is a frame-dependent concept.  To this extent, Bohm and Hiley
suffer from confusions about instantaneity similar to those I have
criticized elsewhere in this paper. Rather, the risk of closed-loop
paradox has to do with the invariant fact that points in spacetime
can sometimes be connected in a closed loop by means of the presumed
superluminal influences; the problem, if any, arises from the fact
that the influences would be superluminal (and thus outside the
light cones of both $A$ and $B$), not that they would be
instantaneous.
%
So the question is whether any putative superluminal theories should
be rejected \emph{just because} they may open up the possibility of
closed causal loops.  I'll return to this point below.

While Bohm's theory is the best-developed causal alternative to
conventional quantum mechanics, it is not the only possible such
theory. Late in his life Louis de Broglie was inspired by Bohm to
revisit his own early attempts at a causal version of quantum
mechanics \cite{deBroglie60,deBroglie70}. De Broglie's late causal
theory, though incomplete in many respects (for instance, it applies
only to spin-0 particles), is fully Lorentz-covariant.
Bohm and Hiley themselves were not comfortable with theories like de
Broglie's later approach (see \cite[p.\ 238]{BH93}) because such
theories imply that any particle interacts \textit{via} the
four-dimensional wave field with other particles both past and
future throughout spacetime. Bohm and Hiley seem to have thought
that this was simply too strong a violation of classical intuitions
about causality.  Thus, Bohm and Hiley rejected covariant pictures
of nonlocality (such as de Broglie's) \emph{not} because they are
technically out of the question, but because they tend to violate
classical expectations or intuitions about causality.

The need to revise our intuitions about causality could be the price
to be paid for any causal interpretation of quantum mechanics that
satisfies the Principle of Relativity.  In particular, a
four-dimensional picture of the wave field could be the answer to
worries about paradoxical closed causal loops: if such loops are
mediated by a genuinely covariant quantum field then it simply would
not be possible to write a description of a self-contradictory loop
in the language of the theory, any more than any other sort of
quantum state vector can be validly written in manifestly
contradictory terms.  That is, while there may well be amplitudes
for past-future-past loops, each possible amplitude could only be
for sequences of events (more precisely, measurement outcomes) that
are mutually consistent. Thus, while such a theory such as de
Broglie's would certainly do violence to classical intuitions
(prejudices?) about the proper order of cause and effect it is quite
likely that it would \emph{not} allow for outright logical paradoxes of the
kind that worried Bohm and Hiley.

A similar picture arises in the two-vector formalism studied by
Yakir Aharonov and collaborators \cite{APT10}. Their theory is not
explicitly a causal interpretation of quantum mechanics, but it also
considers amplitudes from both the past and the future. It could be
worthwhile to investigate parallels between de Broglie's Lorentz
covariant causal theory and the two-vector formalism.  Although
there are closed loops in the two-vector formalism, there is no risk
of paradox for the reason outlined above:  no single looped amplitude is,
in itself, inconsistent.  Like the possible states of
Schr{\"o}dinger's cat, the possible classical outcomes may well be
inconsistent with each other, but each possible outcome set is
internally consistent---\emph{and only one is ever observed}.  In
versions of quantum mechanics that allow for future-to-past
amplitudes, the mystery of causal looping is therefore subsumed into
the larger mystery of understanding the relation between the quantum
mechanical descriptions of physics in terms of amplitudes and the
outcomes that are actually observed. These possibilities require
much further study, but enough is known now to show that one should
not \emph{automatically} reject a version of quantum mechanics
because it allows for causal loops.\footnote{\srr There is a large
literature exploring the puzzle of closed causal loops that could
arise given the possibility of time travel or backwards causation
(not necessarily in the context of quantum mechanics).  Some notable
papers in this genre include \cite{Arnt90,BB92,SmithNJJ97}.  The
upshot of these investigations is that it is by no means
\emph{obvious} that a physical theory should be automatically
excluded because it allows for the possibility of causal looping. }

There is a larger question:  Lorentz invariance itself fails to
satisfy the Principle of Relativity in a certain crucial respect,
since the Lorentz transformations are divergent at a critical
velocity (the velocity of light in vacuum). (Sutherland and
Shepanski \cite{SS86} point to this as the key factor hindering the
extension of the principle of relativity to all relative velocities,
since it makes it impossible to cover all of spacetime, both inside
and outside the light cone, with a single group of continuous transformations.)
It is thus impossible to transform \emph{to} a frame moving with
velocity $c$ and this fact arguably violates the presumption of the
equivalence of all frames. It is conceivable that a deeper theory
which avoids this problem (possibly by allowing for quantum effects
which would suppress the divergence at velocities very close to $c$)
will obey some invariance principle more general than Lorentz
invariance. Let us call such a to-be-written principle \emph{Planck
covariance}. Presumably it would would reduce to Lorentz invariance
in suitable limits just as Lorentz-covariant theories reduce to
Galilean theories in the limit of low relative velocities.

I have indulged in some reasonably well-founded speculations in this
section. However, what is not speculative is that (as the example of
de Broglie's theory shows) it is not necessarily the case that any
account of quantum mechanics in terms of some more general physical
principles would demand the return to Galilean covariance and a
preferred frame; rather, the move to a fully quantized theory of
relativity will probably take us even farther from Galilean
covariance than does special relativity.

\section{Summary---and what must lie ahead}
A lot more needs to be said before anyone has any business being
entirely comfortable with the notion of superluminal influences,
quantum mechanical or otherwise.\footnote{\srr Further problems with
superluminality include but are not necessarily limited to the
following:
\begin{itemize}
\item The temporal order of spacelike separate events is
frame-dependent; this may require the abandonment of causal order as
a global invariant.
\item With some combinations of relative velocities, superluminal
trajectories can form closed causal loops, apparently allowing for
logical paradoxes.
\item Rest mass diverges at $v=c$, apparently precluding the
acceleration of massive bodies through the speed of light.
\item There are problems with reconciling superluminal motion with
local quantum field theory as it is presently understood.
\item In some but not all versions of superluminal or `extended' relativity
proper quantities are imaginary.
\item It is widely though controversially held that quantum
mechanical entanglement cannot be exploited for controllable
superluminal signalling.  (For the orthodox view of quantum
signalling, see, e.g., \cite{ER89,Shimony83}. For critical responses
to this orthodoxy, see \cite{Peacock92,Kennedy95,Mitt98}).
\item The existence of space-like influences or connections demands
a rethinking of the postulate of microcausality which is one of the
building blocks of local quantum field theory.
\end{itemize}

There are candidate responses to all of these problems but they
require discussion that would go far beyond the issues considered in
this paper, which are prerequisites for those discussions. }
But a necessary prerequisite to the analysis of any of the
substantial problems with superluminality is to grasp the kinematics
of propagation outside the light cone.

The following points are elementary even though they have been
persistently misunderstood by professionals working in this field:
\begin{itemize}
\item Trajectories outside the light cone have a natural description in the kinematics of special
relativity.
\item Infinite velocity (equivalently, instantaneity) is a
frame-dependent concept, and thus any form of superluminal propagation is
instantaneous in one and only one frame.
\item The mere existence of some form of superluminal propagation,
even if it is controllable, does not imply the existence, much less
the detectability, of any suppositious absolute state of motion.
\end{itemize}

It is perhaps less immediately obvious, but still clear enough, that
the possibility of distant clock synchronization \emph{via}
superluminal influences does not invalidate the frame-dependence of
time-coordinate simultaneity---because the latter is simply not
about what one could do with superluminal signals, if such things
could exist at all.  And finally it is arguable, though not
conclusively at this stage, that the increasing evidence of dynamic
inseparability in a wide variety of quantum mechanical experiments
(such as the recent dramatic results by Lee \textit{et al.}
\cite{LeeKC11}) points to the cogency of notions of invariant
simultaneity-like relations between spacelike separate entities (or
portions of entities) that are much in the spirit of the ancient
notion of simultaneity as a kind of jointness, wholeness, or
coherence of possibly spatially-extensive events.  The task
remaining is to articulate these possibilities in a precise and
testable way.

\section*{Acknowledgements}
I am grateful to the following people for valuable discussion or advice
about this paper or the topics of which it treats: Richard Arthur,
Bryson Brown, Ad{\'a}n Cabello, Sheldon Chow, Robert Clifton, Saurya Das, Alexander
Korolev, Pamela Lindsay, Nicholas Maxwell, David McDonald, Fred Muller, Vesselin
Petkov, and two anonymous referees for this journal. I am especially
grateful to J.\ R.\ Brown for guidance in the early stages of this
research. Needless to say (but it must be said), none of these
individuals are responsible for any errors on my part in the present
work, and it should not be presumed that they accept my views. For 
financial support I thank the Universities of Toronto,
Western Ontario, and Lethbridge, and the Social Sciences and
Humanities Research Council of Canada. Thanks also to Evan Peacock
for the figures.

\newpage

%
{\frenchspacing\raggedright

}

\end{document}